\documentclass[conference]{IEEEtran}
\IEEEoverridecommandlockouts
\usepackage{cite}
\usepackage{amsmath,amssymb,amsfonts}
\usepackage{algorithmic}
\usepackage{graphicx}
\usepackage{textcomp}
\usepackage{xcolor}
\usepackage[export]{adjustbox}[2011/08/13]
\usepackage{url}
\usepackage{soul,color,graphicx,float,epstopdf}
\usepackage{tablefootnote,textcomp,gensymb}
\usepackage{eurosym,booktabs,multirow}
\usepackage[caption=false]{subfig}
\usepackage{amsfonts} 
\usepackage{algorithmic}
\usepackage{array}
\usepackage{stfloats}
\usepackage{float}
\usepackage{mathtools}
\usepackage{graphicx}
\usepackage{pdfpages}
\usepackage{subfig} 
\usepackage{balance}
\usepackage{xcolor}
\usepackage{comment}
\usepackage[normalem]{ulem}
\usepackage{dsfont}
\usepackage[linesnumbered, ruled]{algorithm2e}
\usepackage{enumitem}
\usepackage{tikz}
\usepackage{scalerel}
\usepackage{tikz}
\usetikzlibrary{svg.path}

\definecolor{orcidlogocol}{HTML}{A6CE39}
\tikzset{
  orcidlogo/.pic={
    \fill[orcidlogocol] svg{M256,128c0,70.7-57.3,128-128,128C57.3,256,0,198.7,0,128C0,57.3,57.3,0,128,0C198.7,0,256,57.3,256,128z};
    \fill[white] svg{M86.3,186.2H70.9V79.1h15.4v48.4V186.2z}
                 svg{M108.9,79.1h41.6c39.6,0,57,28.3,57,53.6c0,27.5-21.5,53.6-56.8,53.6h-41.8V79.1z M124.3,172.4h24.5c34.9,0,42.9-26.5,42.9-39.7c0-21.5-13.7-39.7-43.7-39.7h-23.7V172.4z}
                 svg{M88.7,56.8c0,5.5-4.5,10.1-10.1,10.1c-5.6,0-10.1-4.6-10.1-10.1c0-5.6,4.5-10.1,10.1-10.1C84.2,46.7,88.7,51.3,88.7,56.8z};
  }
}

\newcommand\orcidicon[1]{\href{https://orcid.org/#1}{\mbox{\scalerel*{
\begin{tikzpicture}[yscale=-1,transform shape]
\pic{orcidlogo};
\end{tikzpicture}
}{|}}}}

\usepackage{hyperref} 

\newcommand{\xw}[1]{\textcolor{black}{#1}}

\newcommand{\he}[1]{\textcolor{black}{#1}}

\def\BibTeX{{\rm B\kern-.05em{\sc i\kern-.025em b}\kern-.08em
    T\kern-.1667em\lower.7ex\hbox{E}\kern-.125emX}}
\begin{document}

\title{xApp Distillation: AI-based Conflict \\ Mitigation in B5G O-RAN
\thanks{}
}

\author{\IEEEauthorblockN{{Hakan Erdol \IEEEauthorrefmark{1} \orcidicon{0000-0003-0646-284X}},
{Xiaoyang Wang \IEEEauthorrefmark{2} \orcidicon{0000-0001-9332-2700} },
{Robert Piechocki \IEEEauthorrefmark{1} \orcidicon{0000-0002-4879-1206}}, 
{George Oikonomou \IEEEauthorrefmark{1} \orcidicon{0000-0002-1684-6989}},
Arjun Parekh \IEEEauthorrefmark{3} 
}\\ 
\IEEEauthorblockA{\IEEEauthorrefmark{1} University of Bristol, UK;
\IEEEauthorrefmark{2} University of Exeter, UK;
\IEEEauthorrefmark{3} Applied  Research, BT, UK \\
Email: {\{hakan.erdol,  r.j.piechocki, g.oikonomou \}@bristol.ac.uk}; {x.wang7@exeter.ac.uk} \\
{arjun.parekh@bt.com}
}}
\maketitle

\begin{abstract}
The advancements of machine learning-based (ML) decision-making algorithms created various research and industrial opportunities. One of these areas is ML-based near-real-time network management applications (xApps) in Open-Radio Access Network (O-RAN). Normally, xApps are designed solely for the desired objectives, and fine-tuned for deployment.
However, telecommunication companies can employ multiple xApps and deploy them in overlapping areas. \xw{Consider the different design objectives of xApps, the deployment might cause conflicts}.
To prevent such conflicts, we proposed the xApp distillation method that distills knowledge from multiple xApps, then uses this knowledge to train a single model that has retained the capabilities of previous xApps.  
Performance evaluations show that compared conflict mitigation schemes can cause up to six times more network outages than xApp distillation in some cases.
\end{abstract}


\section{Introduction}
Open radio access networks (O-RAN) in 5G and beyond led to a number of new research and implementation opportunities for the telecommunication sector. One of the key features of O-RAN is xApp applications on beyond the fifth generation (B5G) cellular networks, and the capacity to host machine learning operations. xApp is a part of the RAN intelligent controller (RIC) that takes near-real-time actions as a third-party application to control the network parameters for the desired objectives. The RIC platform has the capability of storing and running multiple xApps \cite{Bonati2021xApp}. ML-based xApps and their capabilities have been studied and proved to be beneficial for next-generation (NextG) cellular networks \cite{MLxApp20211}. However, most of the studies assume there can be only a single xApp in the wireless communication environment and ignore the risk of conflicts \cite{MLxApp20221}. \xw{For sequential decision-making applications, deep reinforcement learning (DRL)-based solutions are among the popular choices for ML-based xApps \cite{Polese2023ColoRAN}.}

O-RAN alliance defined several types of conflict situations\cite{ORANWG3:2021}. 1) Direct conflicts are easy to detect due to the observability of the conflict before actions. Direct conflicts occur when multiple xApps try to control the same parameter in the network. 2) Indirect conflicts cannot be detected until actions are taken in the environment. These situations occur when different xApps adjust different parameters to optimise the same metric according to the respective objective. As an example of indirect conflicts, one xApp manages the resource block allocation while the other controls transmit power. Both of them have effects on the throughput value, but it is not always possible to predict how those actions will affect the performance. Besides the number of potential reasons, effects of implicit conflicts can only be observed on future KPIs which is not possible to avoid at the time of action \cite{ORANWG3:2021}. 


%

To cope with conflicting xApps, the O-RAN Alliance proposes two different conflict mitigation schemes in the RIC architecture technical reports \cite{ORANWG3:2021}. Since the issue is having multiple sets of actions for a single operation, conflict mitigation schemes aim to eliminate all actions and apply the best action among the set. \xw{O-RAN conflict mitigation decides on the optimised action, and disables all of the xApp actions except for the optimised one. This approach doesn't use the information of disabled actions.}

To improve joint coordination of network, \cite{Zhang2022} proposed team-learning. Their approach proposes a joint learning scheme for multiple xApps by sharing actions with each other. Their results show improvements over individually trained DRL-based models in terms of overall throughput and packet drop rates. However, in their scenario, there are only indirect conflicts, and xApps perform only single operations. Moreover, the O-RAN conflict mitigation procedure \cite{ORANWG3:2021} is not applied after RL agents take indirectly conflicting actions.

In this paper, we propose \textit{xApp distillation}, a conflict mitigation scheme that learns from conflicting xApps to create an enhanced xApp with improved performance by using policy distillation.
The major contributions of this letter are summarized as follows:
\begin{itemize}
    \item A novel conflict mitigation scheme that utilises deployed xApps (either ML-based or heuristics) for distilling policy to a better performing, non-conflicting xApp. Current mitigation schemes proposed by O-RAN alliance have shortcomings in service quality. Proposed method provides higher QoS among users.
    \item Proposed xApp distillation forms an xApp that provides computationally efficient and more reliable network management. The method reduces interrupts that are caused by mitigation delay and action rollbacks. 
    \item Proposed method provides a knowledge base by using policy distillation from multiple xApps. Thus, it can be either used to form a comprehensive xApp or an xApp for more specific tasks by the host. 
\end{itemize}

\section{System model}
This paper considers a connected urban system model with multiple users moving in a 250-by-250-meter area. The simulation is built on the \texttt{mobile-env} \cite{schneider2022mobileenv} Gymnasium environment to deploy multiple xApps. We add new resource types and a new reward design to optimise the network. The simulation has \textit{B} base stations (BSs) and \textit{K} mobile users. The designed environment uses an Okumura-Hata-based propagation model \cite{Okumura2000} to simulate an urban area.

\subsection{Connectivity model}
We consider \textit{B} BSs to provide connectivity to \textit{K} users across the area. We defined the downlink data rate $r_{i,j}$ transmitted from BS \textit{j} to user \textit{i} according to calculated signal interference noise ratio (SINR) as follows:

\begin{equation}
\label{eq:datarate}
    r_{i,j} = \sigma R_{i,j} log (1 + \frac{\alpha_{j}  P_{j} h_{i,j}}{N_0 + \sum_{k\neq j} h_{i,k} P_k})
\end{equation}

where
\begin{itemize}
    \item $\sigma $ is the bandwidth of a single RB,
    \item \textit{$R_{i,j}$} is the number of RBs allocated from BS \textit{j} to user \textit{i},
    \item $\alpha_j$ is a binary indicator of connection of BS \textit{j},
    \item \textit{$P_{j}$} is the transmit power level of BS \textit{j}, 
    \item \textit{$h_{i,j}$} is the channel coefficient between BS \textit{j} and user \textit{i}.
\end{itemize}

\subsection{Resource management and xApp design}

The designed environment enables RL agents to manage the network by controlling three type of resources in the network. These network operations in the environment are handover, RB allocation and cell transmit power control.  

\subsubsection{Handover}

We consider \textit{K} users to be mobile in a rectangular area. Due to the change in received signal strength, the controller handovers users from one BS to another to seek better service availability. 

\subsubsection{Resource block allocation}

RB allocation represents the number of RB that is being used by the corresponding user. Although the number of available RBs in LTE was constant, in 5G new radio (NR) it varies depending on the guard bandwidth and single RB bandwidth. We considered using parameters given in Table \ref{tab:5GNR} \cite{Liberg20205Gbook}.
\begin{table}[ht]
\centering
\caption{5G NR resource scheme}
\label{tab:5GNR}
\begin{tabular}{l|l}
\hline
Channel bandwidth        & 100 MHz \\ \hline
Guard bandwidth          & 845 KHz \\ \hline
One RB bandwidth         & 360 KHz   \\ \hline
Total number of resource blocks & 273   \\
\hline
\end{tabular}
\end{table}

As a typical profile mode of 5G, we consider one RB to be 360 kHz and with the double guard band, there are 273 RBs in total.

\subsubsection{Transmit power level control}

Each BS in the environment has various transmit power levels. The power levels directly affect the datarate, hence it creates an indirect conflict possibility with RB allocation actions. Moreover, we used SINR values instead of SNRs to penalise RL agents which frequently maximise the transmit power level. In the simulation, we consider changes in cell power level between 25 dBm to 35 dBm.

\section{xApp Distillation}

The conflict mitigation schemes proposed by O-RAN ignore some of the xApps' actions to mitigate conflict by choosing one of the options and discarding others. Therefore, the chance to take a sub-optimal action in the environment becomes higher. Hence, we proposed the xApp distillation method that utilises all xApps by distilling their policies to form a single xApp that consists of all action capabilities of the candidate xApps. 
We build a knowledge base using experiences from all xApps's transition memories. These memories are collected in a replay buffer memory to train new xApps. Policy distilling is applied by having a teacher network i.e., pre-trained xApps to train a new model for the student network i.e., new xApp.

\subsection{Conflicting xApps}

In this paper, we designed two xApps that have multiple operation capabilities. In xApp 1, we developed handover and RB allocation while xApp 2 operates handover and transmit power level control of BSs. The objective of this design is to introduce both direct and indirect conflicts simultaneously. Hence, we could analyse how xApp distillation conducts the mitigation for these xApps.

\subsection{Deep Q-Network}

As a network controller we employed deep Q-network because of its advantages over tabular Q-learning on continuous state spaces. In the simulated environment according to users' trajectories, there can be an indefinite number of state-action transitions. Therefore, using a tabular-based method may be less likely to converge to an optimal solution.

In this paper, we designed a multi-headed multilayer perceptron (MLP) network model for RL agents. Therefore, each DQN model can perform multiple tasks in a single production of inference. For the xApps each head of MLP model corresponds to an operation such as a handover for users. The main reason to choose a multi-headed network structure is to avoid a sequential decision making process for users. Because in sequential decisions first users tend to take greedy actions while last ones are left with suboptimal options due to BS capacity or depleted RBs at cells.

\subsubsection{State}

State includes BS connection indicator, SINR values for neighbouring BSs and a utility parameter for each user. The connection indicator is a one-hot encoded array of BS connections $\alpha = [\alpha_1 , ... ,\alpha_N ]$. 


SINR value for user \textit{i} is calculated according to their distances to BSs. For BS j, it is calculated as follows:

\begin{equation}
    SINR_j = \frac{P_j h_{i,j}}{N_0 + \sum_{k\neq j} h_{i,k} P_k}
\end{equation}

The last parameter of the state information is the utility value of users. Utility parameter provides an interface between state and reward to fulfill the Markov decision process (MDP). Utility corresponds to the logarithmic representation of downlink data rate $U_i = 10 [\frac{ln(r_i)}{ln(w)}]_{L}^{U}$.


where \textit{$r_i$} is the data rate, \textit{w} is a coefficient to scale the utility parameter. U and L are the upper and lower limit respectively. The overall state vector is given as follows:

\begin{equation}
    State = [\alpha , SINR_{j=1:B} , U_{i=1:K}]
\end{equation}

\subsubsection{Action}

The action space represents the decisions that xApps produce according to the state of the environment. The type of action defines the ability of xApps. There are three different operations namely, handover, RB allocation and cell transmit power control. In this study, we wanted to simulate both direct and indirect conflicts as classified in \cite{ORANWG3:2021} through the actions of RL agents. 

To simulate direct conflict, we created two xApps, both employ handover capability. For the indirect conflict, xApp 1 allocates RBs among the users while xApp 2 controls the cell transmit power in addition to the handover operation. Both the number of allocated RBs and the transmit power level affect the overall channel capacity of users. Since these are different types of actions observing such conflict prior to action is almost impossible \cite{ORANWG3:2021}.

In the simulation, there are three types of actions and each xApp has a different ability to manage the network. 

\begin{itemize}
    \item xApp 1: Handover and RB allocation
    \item xApp 2: Handover and cell power level control
    \item Final xApp: Handover, RB allocation and cell power level control
\end{itemize}

The action vectors of DQN models are designed as follows:

\begin{equation}
\begin{aligned}
    \alpha &= [[DC, \alpha_{1,1}, ... , \alpha_{1,B}],..., [DC, \alpha_{K,1}],...,[\alpha_{K,B}]] \\
    R &= [[R_{1,1},..., R_{1,M_R}] , ... , [R_{K,1}, ... ,R_{K,M_R}]] \\ 
    T &= [[PL_{1,1}, ... , PL_{1,M_T}],...,[PL_{B,1}],...,[PL_{B,M_T}]] \\ 
    Action &= [ \alpha, R, T]
\end{aligned}
\end{equation}

where \textit{R} indicates the number of RB assigned to \textit{K} users up to \textit{$M_R$} blocks, \textit{T} is the transmission power levels for \textit{B} BSs up to \textit{$M_T$} level. We added "\textit{DC}" disconnect action as a handover option to analyse how often RL agents cannot find an optimal BS to connect. Each operation in the action vector corresponds to the output head of the designed multi-headed DQN network.

\subsubsection{Reward}

Reward design plays a crucial role in DRL applications that train a model to perform actions for specific tasks. Because it defines the task to which the model is trained to perform.

To avoid greedy actions for specific users and poor performance for the rest of the users, we chose to use proportional fairness as the reward function. While it increases with the throughput, it penalises too high or too low data rates by using a logarithmic data rate. The reward function for \textit{K} users is calculated as follows:
\begin{equation}
    PF = \sum_{i=1}^{K} log(r_{i})
\end{equation}

For a fair service availability for each user, we employed proportional fairness where downlink data rate \textit{$r_{i}$} is calculated as in equation \ref{eq:datarate}. 
\subsection{Policy distillation}

\begin{figure}[htbp]

    \centerline{\includegraphics[width=0.4\textwidth]{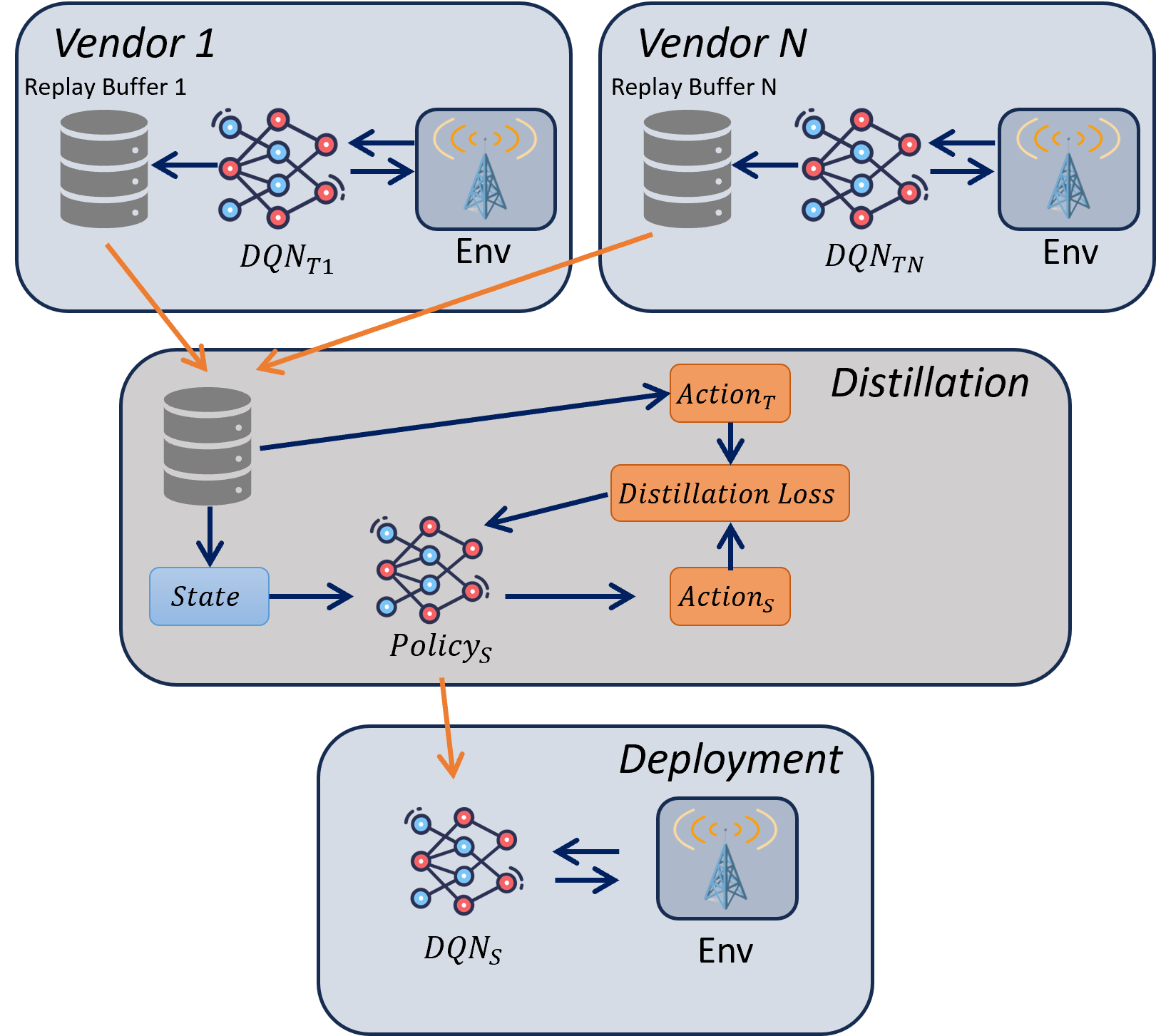}}
    \caption{O-RAN Architecture in the proposed system model.}
    \label{fig:Arch}
\end{figure}

Policy distillation, originally proposed for transferring a pre-trained action policy into an untrained Q-Network \cite{rusu2016policy}. In this paper, we consider a scenario in which the host purchases multiple xApps from different vendors and deploys them into the network that provides service on partially or fully overlapping areas. Since the host cannot intervene with the training stage of those xApps, our proposed method uses these pre-trained xApps to distil knowledge to a student policy that will take action on behalf of all teacher networks. To simplify the process, we divided it into four stages. The stages of xApp distillation are illustrated in figure \ref{fig:Arch}.

To simulate this scenario, we individually train these xApps in their environments until their performances converge to an optimal performance level. This stage i.e., stage 1, corresponds to the phase where different vendors train their xApps in their environments. 

After completing stage 1, we assume that a host purchases these xApps without knowing the model parameters of the DQN models or any training-related features. Then, we consider a scenario where the host deploys these xApps to observe the output (actions) according to the given input (observations of the state). Then, we repopulate a buffer memory by using these pre-trained xApps. Since the deployed environment drastically changes the outcome of RL interactions, we consider deploying pre-trained xApps on the environment in which the final xApp will operate. At this point, pre-trained xApps act as teacher models. Thus, individually deployed xApps generate data to fill a replay buffer memory. \he{Note that we save the information before it is applied to the network. Otherwise, the network controller can change the action when applying due to O-RAN conflict mitigation}. When sufficient state, action pairs are stored in replay buffer, stage 2 ends.

\begin{figure}[htbp]

    \centerline{\includegraphics[width=0.5\textwidth]{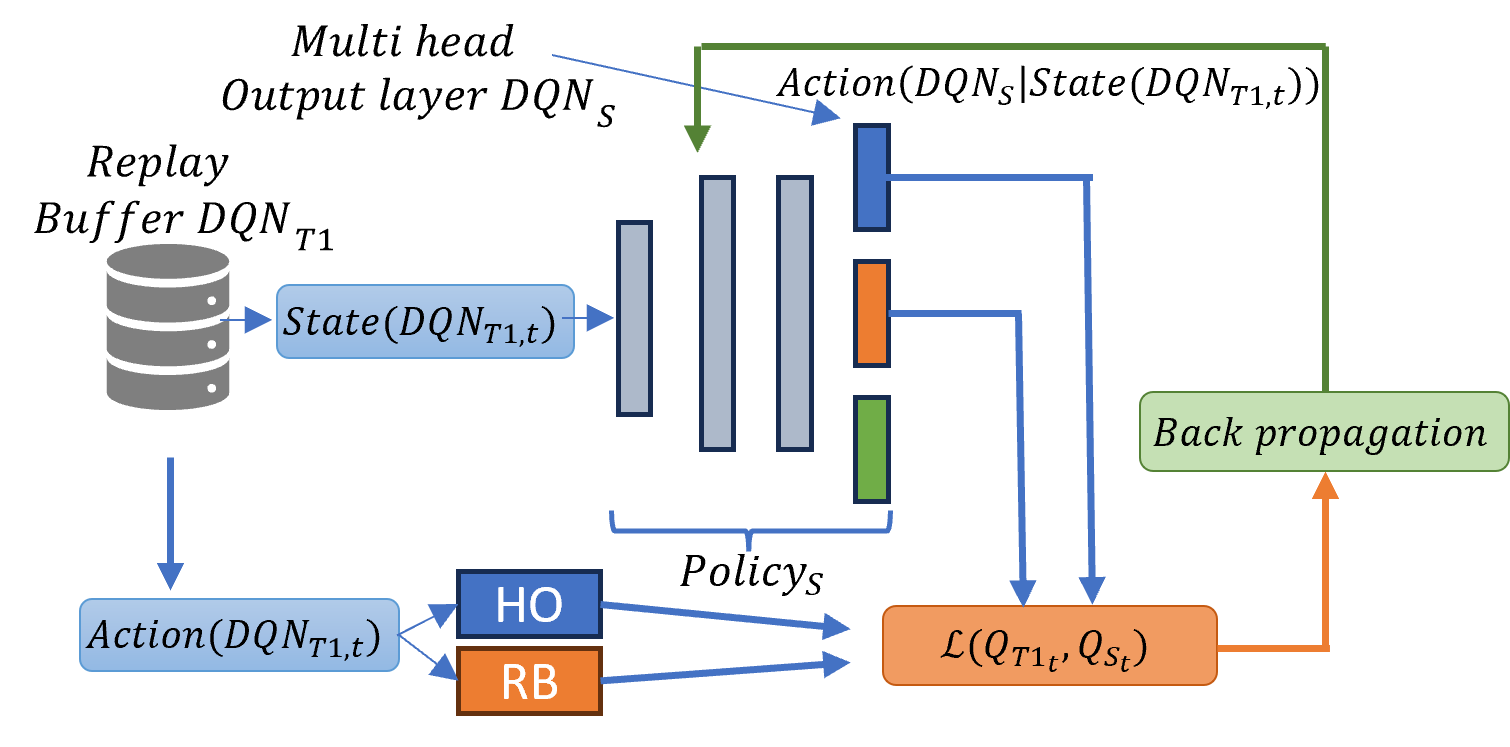}}
    \caption{Stage 3 - Distilling knowledge from pre-trained xApps to a student policy.}
    \label{fig:Dist_Dia}
\end{figure}

In stage 3, we use replay buffer memory to extract state and action data for distillation. xApp distillation is essentially a supervised learning scheme which uses well-trained xApps i.e., teacher model, and trains an untrained deep Q-network i.e., student model. These well-trained xApps do not necessarily have to be an ML-based method. As long as they produce an action regarding to state, they can be heuristics or a mix of both as well. Note that the action space of teacher networks and student policy can be different. As shown in Figure \ref{fig:Dist_Dia} loss is calculated by only using the matching heads of the teacher DQN model and the student policy.

The stage 4 of the xApp distillation is the deployment of the student policy to a DQN model for test and evaluation. In our simulations we experimented with distilling policies from heuristics, however distilling policy from another DQN model performed better. 

%

When distilling policy from a teacher network, choosing distillation loss plays a crucial role in the process. As proposed in \cite{rusu2016policy} we employed KL divergence loss:


\begin{equation}
    L_{KL} ( \mathcal{D}^T , \theta_S ) = \sum_{i=1}^{|D|} softmax(\frac{q_{i}^{T}}{\tau}) ln \frac{softmax(\frac{q_{i}^{T}}{\tau})}{softmax(q_{i}^{S})}
\end{equation}

where $\mathcal{D}$ is replay buffer memory, $\tau$ is the temperature, $q_{i}^{T}$ and $q_{i}^{S}$ are the Q-values from the teacher and student networks respectively. 

%
%
%
%
%
%
%
%
%
%
%
%
%
%
%
%
%
%
%
%
%

\subsection{Conflict mitigation scheme}

O-RAN conflict mitigation method \cite{ORANWG3:2021} essentially proposes two types of mitigation namely against direct and indirect conflicts. 

For direct conflicts, it is easy to observe and avoid conflicts. O-RAN alliance proposes simply ignoring one of the xApps and applying the other for the network. 

However, for indirect conflicts, it is almost impossible to observe conflicts prior to the action. Therefore, they propose to monitor the network performance, and if the network performance deteriorates after the performed action, it rolls back the action and applies previous actions. These are the essential mitigation scheme for inter-xApp conflicts.

In both cases, it ignores some of the xApps and applies an action taken by different xApps or taken for the previous state of the network which leads to possible sub-optimal actions. 

However, the xApp distillation utilises all xApps' abilities by using their experiences on the policy distillation. Moreover, after xApp distillation, a single xApp which is the result of distillation is deployed on the environment. Thus, it eliminates inter-xApp conflicts. 

\section{Performance Evaluation}
In this paper, we use the "mobile-env" \cite{schneider2022mobileenv} wireless communication environment wrapped in the Gymnasium framework as a base of our simulations. To implement conflicts we added new resource types to control such as resource block allocation among users and base station transmit power level control.  

We compared our results with individually trained and deployed xApps, and xApps that are trained using the team learning method \cite{Zhang2022}. Both methods are subject to O-RAN conflict mitigation due to the deployment of multiple xApps in a single environment.

\subsection{Performance parameters}

We simulated this method using DQN models as the target of distillation. However, this is an open method to other DRL-based controllers. The DQN model structure we used for xApps is relatively lightweight. The structure of the model is an MLP, consisting of 4 layers, and hidden layers have 50 and 100 neurons respectively. We used SGD optimiser to update teacher xApp models throughout 100,000 episodes of pre-training with a learning rate of 0.01. After pre-training was completed we used 10,000 post-training steps to save \textit{$S,A,R$} transitions that will be used to distil knowledge to the student model. The temperature value for distillation is set to 20.

\subsection{Numerical results}

\begin{figure}[htbp]

    \centerline{\includegraphics[width=0.4\textwidth]{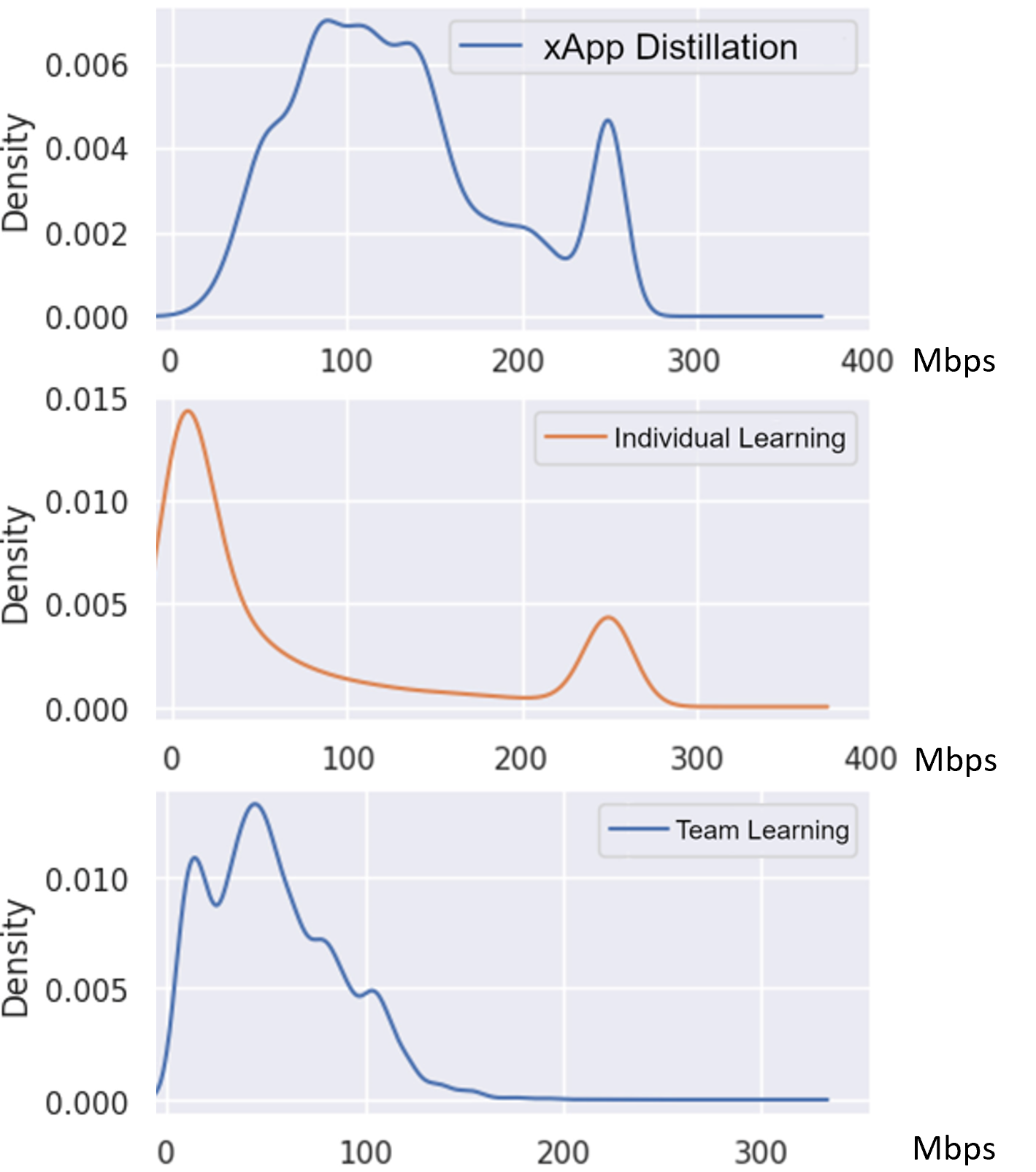}}
    \caption{PDF of throughput comparison between the xApp Distillation method and O-RAN conflict mitigation with models trained either with individual or team-learning scheme.}
    \label{fig:PDFComparison}
\end{figure}

O-RAN conflict mitigation applies one of the xApps during the conflicts and eliminates the rest. Therefore, alternating network control options creates inconsistent service for the users. In Figure \ref{fig:PDFComparison}, we illustrated the probability density function (PDF) of downlink data rate for the users to see the distribution of data-rate to analyse consistency. We compared our results with two different scenario. In Individual learning, the host purchases pre-trained xApps and deploys them in the same area by employing conflict mitigation by O-RAN Alliance. In team learning \cite{Zhang2022}, xApps are trained jointly by receiving feedback on the action of the other xApp.  

The figure compares the xApp distillation with O-RAN conflict mitigation applied to individually trained agents and the agents trained by using the team-learning method \cite{Zhang2022}. While the team-learning method improves the performance of individually trained and deployed xApps, xApp distillation shows overall the highest data rates. 

\begin{figure}[htbp]

    \centerline{\includegraphics[width=0.45\textwidth]{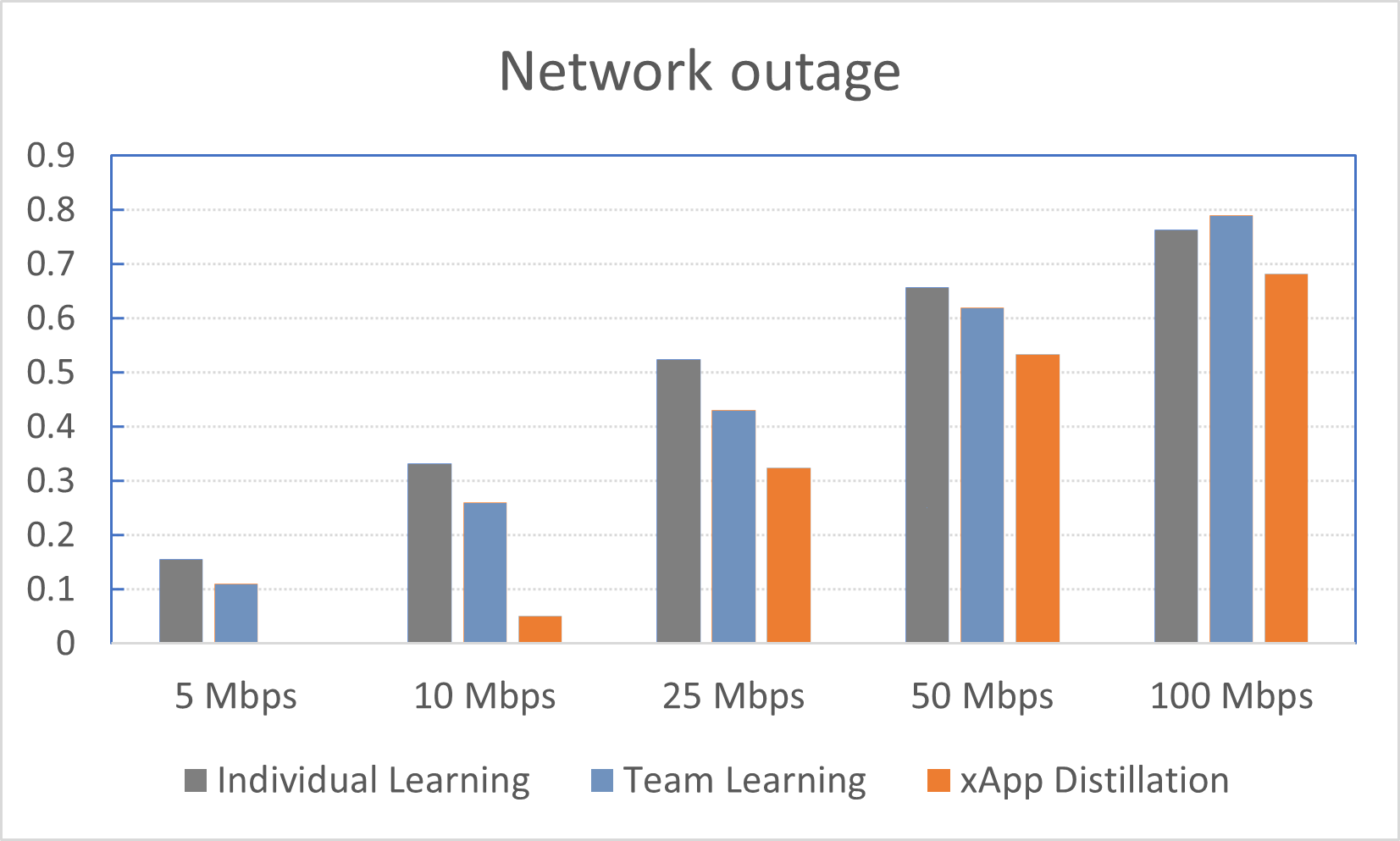}}
    \caption{Network outage comparison in percentage between the xApp Distillation method and O-RAN conflict mitigation with models trained either with the individual or team-learning scheme.}
    \label{fig:NetOutage}
\end{figure}

To evaluate these results, we compared these methods in terms of network outage in 50,000 time steps in the environment. The network outage for 5 Mbps to 100 Mbps for all methods is given in Figure \ref{fig:NetOutage}.

The figure demonstrates the network outage ratio for Individual Learning, Team Learning, and our xApp Distillation method across different bandwidths. Notably, xApp Distillation significantly outperforms the other methods. For instance, at 10 Mbps, it achieves an 83.3\% reduction in network outage compared to Team Learning. Even at higher bandwidths like 50 Mbps, xApp Distillation maintains a 33.3\% improvement over Individual Learning. This highlights xApp Distillation's effectiveness in consistently reducing network outages and enhancing reliability.

\section{Conclusion}
In this paper, we show that xApp distillation eliminates inter-xApp conflicts and it provides significantly better performance than state-of-the-art studies in terms of reliable and consistent service. Since indirect and implicit conflicts are almost impossible to detect or mitigate, our method takes advantage of using the experience of pre-trained xApps and take actions from a single decision-maker instead of applying multiple xApps actions simultaneously which can cause such conflicts. xApp distillation provided a consistent 10 Mbps downlink data rate where compared methods failed to provide up to more than 30 percent simulation runtime. 

In our future work, we plan to extend this xApp distillation method to use more scalable RL controllers than DQN method such as Multi-agent RL  and decomposed action-space methods. Moreover, we will increase the number of KPIs to analyse network in a more comprehensive way.

\section*{Acknowledgment}

This work has been funded by the Next-Generation Converged Digital Infrastructure (NG-CDI) Project, supported by Engineering and Physical Sciences Research Council (EPSRC), Grant ref. EP/R004935/1.

The author, Hakan Erdol, is supported by a scholarship from the Turkish Ministry of National Education during this study.

\bibliographystyle{IEEEtran} 

\bibliography{IEEEabrv,references}

\vspace{12pt}

\end{document}